# Recovery of Meteorites Using an Autonomous Drone and Machine Learning

Robert I. Citron[1], Peter Jenniskens[2], Christopher Watkins[3], Sravanthi Sinha[4], Amar Shah[5], Chedy Raissi[6], Hadrien Devillepoix[7], and Jim Albers[2]

[1]*Department of Earth and Planetary Sciences, University of California, Davis 95616, USA.*
[2]*SETI Institute, Carl Sagan Center, Mountain View, CA 94043, USA; NASA Ames Research Center, Moffett Field, CA 94035, USA.*
[3]*Commonwealth Scientific and Industrial Research Organisation, Scientific Computing, Clayton, VIC 3800, Australia.*
[4]*Holberton School of Software Engineering, San Francisco, CA 94111, USA.*
[5]*Cambridge University, Dep. of Engineering, Computational and Biological Learning, Cambridge, CB2 1PZ, UK.*
[6]*Institut National de Recherche en Informatique et en Automatique, France.*
[7]*Space Science Technology Centre, School of Earth and Planetary Sciences, Curtin University, GPO Box U1987, Perth, Western Australia 6845, Australia.*

June 11, 2021

**Abstract**

The recovery of freshly fallen meteorites from tracked and triangulated meteors is critical to determining their source asteroid families. However, locating meteorite fragments in strewn fields remains a challenge with very few meteorites being recovered from the meteors triangulated in past and ongoing meteor camera networks. We examined if locating meteorites can be automated using machine learning and an autonomous drone. Drones can be programmed to fly a grid search pattern and take systematic pictures of the ground over a large survey area. Those images can be analyzed using a machine learning classifier to identify meteorites in the field among many other features. Here, we describe a proof-of-concept meteorite classifier that deploys off-line a combination of different convolution neural networks to recognize meteorites from images taken by drones in the field. The system was implemented in a conceptual drone setup and tested in the suspected strewn field of a recent meteorite fall near Walker Lake, Nevada.

# 1 Introduction

There is an ongoing effort to determine the composition of some 40 asteroid families in the asteroid belt to understand the early evolution of the solar system. Remote sensing information is complemented with data gleaned from laboratory studies of meteorites. To put that meteoritic data in context, the approach orbit of freshly fallen meteorites are measured to determine what asteroid family might have produced the meteoritic debris from a particular collision event. This is a statistical effort to determine the inclination and semi-major axis distribution of the approach orbits of meteorites of the same type with the same collision age. The inclination identifies the inclination of the source region, while the semi-major axis points to the delivery resonance (Jenniskens, 2013).

If the meteorite can be recovered, a fireball's lightcurve and deceleration profile also provides information about how its kinetic energy is deposited in the Earth's atmosphere, which is a function of density and internal strength. That information can be used to improve predictions at what altitude asteroids of this material type fragment that are big enough to cause damaging airbursts (Popova et al., 2013).

So far, in only about 40 cases have meteorites been recovered from observed falls for which the fireball trajectory in the atmosphere and pre-impact orbit were measured (Borovička et al., 2015; Jenniskens, 2013).

Many of those are cases where the meteorite was found after which the fireball data was obtained from serendipitous security or dash-cam footage. Until now, various fireball networks have recorded approximately 800 trajectories of meteoroids significant enough to have dropped meteorites on the ground, of which only $\sim$26 cases (3%) resulted in meteorite recovery.

Finding meteorites from an observed fall can be very difficult. Meteorites are scattered over many square kilometer large strewn fields and typically are found only through a physical survey, taking $\sim$ 100 man-hours to locate one meteorite fragment. The investment of that time is made more often when the falling meteorites are detected by Doppler weather radar, as it gives confidence that meteorites survived to the ground (*e.g.*, Jenniskens et al., 2012).

In order to increase the recovery yield of the more frequent smaller falls, a more efficient method of searching for freshly fallen meteorites is required. Recent advances in autonomous drone technology and machine learning image classification algorithms have enabled the possibility for autonomous meteorite fragment localization. Autonomous quadcopter drones can navigate terrain at a fixed height, obtaining a survey of top-down images. These images can be spliced and fed into a machine learning object detection classifier, which can determine the likelihood that meteorite fragments are present in the image. Using a machine learning object detection model coupled with an autonomous drone, it may be possible to reduce the man-hours needed to locate freshly fallen meteorites, and increase the likelihood of locating meteorites in strewn fields.

## 1.1 Prior work

The use of drones surveys to locate freshly fallen meteorite fragments has gained increasing traction in recent years (Citron et al., 2017; Zender et al., 2018; AlOwais et al., 2019; Anderson et al., 2020). The general method of detecting meteorites with a drone survey involves applying a machine learning classifier to images taken from an autonomous drone that first surveys the area at low altitude (Citron et al., 2017). With a sufficient field of view and image resolution, each 1-3 cm meteorite fragment should be resolved by >20 pixels in width, sufficient for image classification. If a machine learning classifier can be effectively trained to locate meteorites in drone-acquired images, a survey of aerial photos can be spliced and classified to determine if any meteorite fragments are present.

Training a machine learning classifier to detect freshly fallen meteorite fragments presets a challenge. The main identifying characteristic of a meteorite fragment is that it should appear to not belong in the surrounding terrain. Although freshly fallen meteorite fragments should appear quite distinct from native rocks, meteorites are diverse and it is impossible to know the characteristics of the next freshly fallen meteorite fragments or the surrounding terrain. While meteorites typically are darker than native rocks due to the fusion crust obtained during atmospheric entry, some meteorite fragments lack a fusion crust because of subsequent break-up. The challenge is also that meteorite strewn fields can result in search areas several square kilometers. In such a search area, there may be only a few meteorite fragments 1-3 cm in size. A machine learning classifier therefore must be versatile enough to detect meteorites of a variety of types on completely new terrains, and return sufficiently low false positive detections to make a square kilometer survey feasible.

The construction of a machine learning classifier to identify meteorite fragments from native terrain should be feasible, and several groups have tested the application of machine learning to meteorite detection in the field (Citron et al., 2017; AlOwais et al., 2019; Anderson et al., 2019; Anderson et al., 2020). The most suitable machine learning classifier for meteorite identification might be based on convolutional neural networks (LeCun et al., 1998), which have achieved state of the art performance in a variety of computer vision based tasks (Gu et al., 2018). A classifier based on convolution neural networks avoids problems associated with hand picking features, which would be inadequate for meteorite identification because of the diverse possible properties of meteorite fragments. The key benefit of a neural network based classifier is that it is trained to learn the features most relevant for its task jointly with the output predictions it makes. Thus, by applying deep learning algorithms to the task of discriminating between images of patches of land with and without meteorites, a general classifier can be constructed to identify a freshly fallen meteorite fragment on most terrains.

There are two main types of machine learning models applied to object recognition: binary image classifiers or object detection networks. A binary image classifier utilizes the entire image to determine what object it represents. Binary image classifiers are trained on sets of 'positive' and 'negative' images, and when a trained binary classification network is applied to a new image it determines the likelihood the image falls in the 'positive' or 'negative' category (*i.e.*, if the image is of a meteorite or not). For this type of classifier to work the meteorite must encompass most of the image, therefore the network must be applied to small image patches. A large image acquired by a drone must therefore be spliced into multiple smaller image patches before being fed into the classifier (Citron et al., 2017). Alternatively, an object detection network may be used. Object detection networks are trained on sets of larger images where each class of object is demarcated (usually by a box) within each training image. Instead of training on separate sets of positive and negative images, the object detection network learns to recognize classes of objects (or a single class of objects, *e.g.*meteorites) marked within each image as 'positives' and the rest of the image where the object class is not highlighted is effectively 'negative'. When a trained object detection network is applied to a new image it outputs a box around each section of the image where it predicts the object classes are located. Object detection networks can be fed an image of any size and locate an object within the image. This is more time efficient because each drone acquired image does not need to be spliced into smaller fragments that need to be classified independently.

Several studies have tested the use of a machine learning classifier to find freshly fallen meteorite fragments. An earlier version of our system was tested in Creston, California (Citron et al., 2017), however, that iteration used binary image classifier that required each image to be spliced into ∼64x64 pixel patches before classification. This resulted in long processing times for each full image. Furthermore, many false positives were detected, depending on the terrain type. Zender et al. (2018) also tested meteorite detection using a drone mounted camera, but did not use a machine learning classifier and instead applied a simple algorithm based on reflectance characteristics. AlOwais et al. (2019) performed real-time detection of meteorites by implementing a machine learning algorithm on board a drone. Their algorithm processed images from a live video feed and achieved an accuracy of ∼ 90%. Anderson et al. (2020) implemented the Keras machine learning module applied to 200x200 pixel tiles spliced from drone-acquired images. They constructed a training dataset by pasting meteorite images into drone-acquired images of the local terrain and trained a binary image classifier to identify meteorite fragments with validation accuracy of ∼97%. In a field test, Anderson et al. (2020) used an RGB camera mounted on a multicopter to survey a searching area, covering 1 km$^2$/day at a resolution of 1.8 mm/pix. Their machine learning approach was able to correctly identify 3 real meteorites (two falls, one find) in their native fall locations.

Machine learning is a rapidly advancing field and is providing new tools to increase the accuracy of object detection and classification. Of particular interest are residual neural networks, which have achieved excellent performance in visual recognition tasks (He et al., 2015). Residual learning framework can be used to train deeper neural networks capable of advanced object detection independent of the image size. Thus, instead of training a binary image classifier to examine many small patches (<200 pixels in width) (*e.g.*, Citron et al., 2017; Anderson et al., 2020), an effective object detection network can be trained to recognize small meteorite fragments in full images. Utilizing these recent advances, we train a residual neural network based object detection network to recognize meteorite fragments in drone acquired images, and test its accuracy in the field.

Here, as a proof-of-concept, we construct an object detection network using images of meteorite fragments collected from an autonomous drone. We chose an inexpensive off-the-shelf drone model (3DR) and camera (GoPro Hero4) in hope that our algorithm could later be applied on a much bigger scale by owners of such equipment. We show that classification of meteorite fragments in drone images is possible, and test our model at the location of a recent suspected strewn field in Walker Lake, Nevada.

## 2 Methods

### 2.1 Constructing the Machine Learning Classifier

#### 2.1.1 Training Dataset

Any machine learning classifier requires a large training data set of thousands of positive and negative images. Because a future meteorite fall could occur on any type of terrain, and consist of any type of meteorite fragment, we constructed a dataset containing a variety of meteorite types on a diverse set of terrains. A classifier trained on such a set of images could be readily applied to the classification of images from a future fresh meteorite fall. We used three sources for meteorite images to construct a large training dataset: 1) placing a limited collection of 8 meteorite fragments in our possession on various local terrains and taking overhead images of them, 2) placing our collection of 8 meteorite fragments in a recent meteorite strewn field and imaging them from above with a drone, and 3) using overhead shots of freshly fallen meteorite fragments obtained from an internet search.

For collecting images of meteorite fragments on local terrains, we used our collection 8 meteorite fragments 10-100 g in mass (1-4 cm in diameter) obtained fresh from the 1992 Mbale meteorite fall Jenniskens et al. (1994). The meteorites were deployed on various local grass, dirt, sand, and rocky terrains, and imaged with smart phone cameras and a DSLR from above at a similar height as the expected drone images. Each image was spliced into smaller patches of 1000x600 pixels. Although the object detection network described in Section 2.2 could work on images of any size, we found that processing was limited by the computer memory requirements during image classification, and 1000x600 pixel image patches resulted in good performance. Examples of image patches containing meteorites used in training the object detection classifier are shown in Figure 1a. We used 526 image patches with meteorites placed on various local terrains and imaged from above.

The second source of test images was from a previous test of an earlier iteration of our machine learning classifier. During this prior test, we deployed our meteorite fragments in the location of a recent meteorite fall near Creston, California and imaged the fragments from above with a drone-mounted camera (Citron et al., 2017). For the updated object detection classifier described in this paper, we used the images from our previous test as an additional data source to train our new classifier. Because these images were taken with the same camera and drone system, they are more applicable to future field deployments and of high value to training our updated model. Each image was spliced into smaller 1000x600 pixel image patches before being added to the training dataset. Example training data from our field test are shown in Figure 1b. We used 82 image patches obtained from a drone-mounted camera during our field test in Creston, CA.

The third source of training images was obtained from an internet search for pictures of freshly fallen meteorites. The use of images of meteorites obtained from the web was critical because we did not want to overtrain our object detection network on the 8 meteorites in our collection. By using images of meteorites from various falls worldwide, we were able to include in our data a variety of different meteorite types and background terrains, enhancing our ability to deploy our machine learning algorithm at the site of a future fresh fall. The meteorite images obtained online were re-scaled or cropped so that each image patch was 1000x600 pixels, to construct a training dataset of similar image sizes. Overall, we used 155 image patches from meteorite images collected from the web. Examples of pictures of meteorites obtained from the web are shown in Figure 1c.

Overall, our training dataset contained 526 images of our 8 fragments placed on local terrains, 82 images of our 8 fragments taken during a field test, and 154 images of meteorites obtained from an internet web search. These image sources were combined into a single dataset of 762 images. The training data set was separated into a training and validation subset. The validation subset was 150 images randomly selected images ($\sim$ 20% of the full dataset) and the training subset was the remaining 612 images. Because the number of meteorite images in our dataset was limited, we augmented the training dataset by reflecting each image across the horizontal plane, vertical plane, and both the horizontal and vertical planes, resulting in a total of 2448 images in the training dataset.

## 2.2 Training the Object Detection network

For meteorite detection, we use the RetinaNet object detection network (Lin et al., 2017). RetinaNet is a highly efficient object detection network that is built upon the ResNet deep residual neural network architecture (He et al., 2015). The implementation of RetinaNet utilizes the Keras (Chollet, 2015), Caffe (Jia et al., 2014), and Tensorflow (Abadi et al., 2015) software packages. We use the ResNet-50 backend (He et al., 2015) which is pre-trained using a dataset of roughly 15 million images called ImageNet (Russakovsky et al., 2014). This approach means that from initializing our training phase, the object detection network would already be familiar with objects such as grass, rocks, hay, etc, and would not have to learn their features from scratch. We took the pre-trained model and trained it on our dataset of meteorite images for several epochs with RetinaNet in order to obtain a final object detection model. We trained the model

a) Samples deployed on local terrains

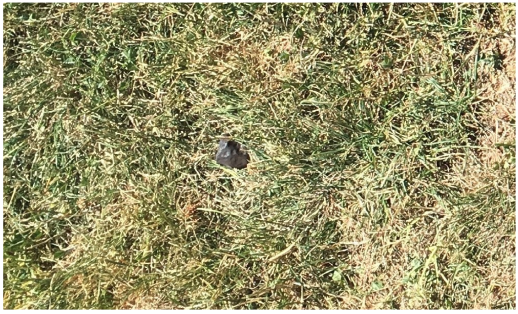 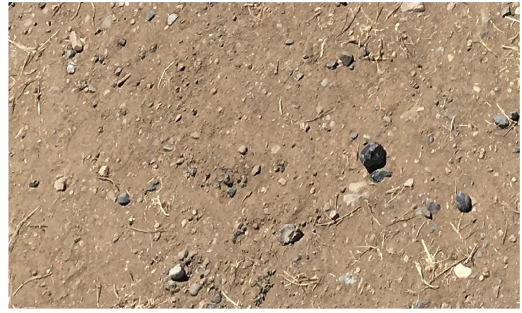

b) Samples imaged with the drone during a field test

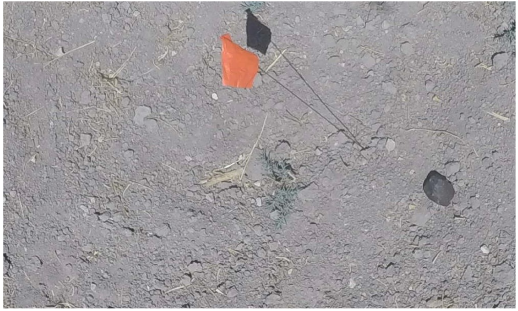 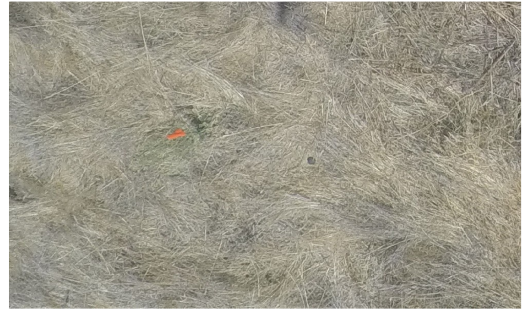

c) Images of freshly fallen meteoites obtained from the web

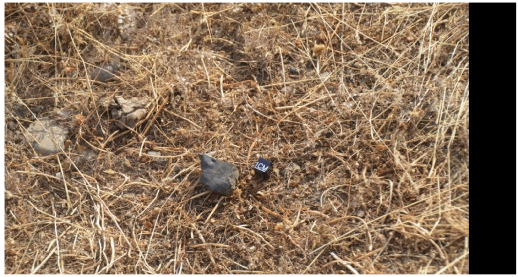 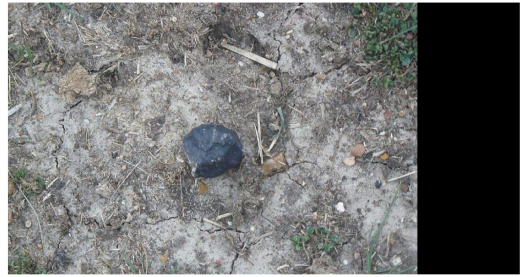

Figure 1: Examples of images used to train our object detection dataset. (a) Images of two meteorites from our collection placed on local terrains and captured from above with a DSLR camera. (b) Images of two moites from our collection placed in a field test in Creston, California and captured from above with a GoPro Hero4 camera mounted on the 3DR Solo Drone. (c) Example images of freshly fallen meteorites obtained from an internet search.

for three epochs until the classification loss was <0.05 to obtain high accuracy without over-fitting to our training data.

Once trained, RetinaNet could process a 1000x600 pixel image in 135−146 ms on our field laptop equipped with an NVIDIA GeForce GTX 980 graphics card. Each image taken with the drone-mounted GoPro camera was 4000x3000 pixels (12 Mp), and was spliced into 47 1000x600 pixel patches using a 200 pixel spacing to avoid splitting any potential meteorites across image patches. This resulted in a total processing time of ∼ 6.6 seconds for each full GoPro image.

While RetinaNet obtained a low classification loss, field deployment will involve identification of many false positives. This is inevitable considering that many other dark rocks might be mistaken for meteorite fragments. It is important to emphasize that the goal of the object detection network is not to return only a few images of suspected meteorites with 100% accuracy within a given search area, but instead return more on the order of ∼ 100s of candidate image patches that might contain a meteorite fragment within the search area. The user can then scan through the image patches and flag particularly strong candidates for follow-up.

## 2.3 Drone hardware and field deployment

The drone currently used in our deployment is the 3DR Solo quadcopter drone fitted with a gimbal-mounted GoPro HERO 4 camera. Because of the hilly terrain in much of California, we upgraded the 3DR Solo drone with a laser altimeter and the PixHawk GreenCube flight controller capable of running Arducopter 3.4. The use of the Arducopter 3.4 equipped flight controller and the laser altimeter allowed the drone to use true terrain following to maintain a constant elevation above the ground. We upgraded the GoPro with a narrow-angle lens offering a 87 degree field of view. Flying at a height of 3 m with the 4000x3000 pixel GoPro camera, this yields a resolution of 0.97 mm/pixel. At 6 m, the resolution is 1.95 mm/pixel. The desired resolution is $\sim$ 20 pixels across a meteorite fragment, so the drone was flown at 3 m when searching for smaller fragments $\sim$ 2 cm across (10-20 g) and 6 m when searching for larger fragments $>$ 4 cm across (75-150 g).

Each drone survey was programmed with the Mission Planner software, which allowed us to pre-fetch the map data prior to field deployment. During field deployment we programmed the drone to fly a grid-search pattern and take photos from a constant altitude of $2-6$ m. Because of the battery-limited $\sim$ 25 minute flight time of the 3DR Solo drone, each grid survey acquired $\sim$ 130 to 250 images, depending on the flight altitude and search pattern. Because our object detection classifier identifies many false meteorite candidates, the user must scan through the processed images to determine which candidates are worth following up with another field visit. Each image is tagged with a timestamp that can be compared to the drone's GPS log to determine the approximate location where the image was acquired. Although in principal it is possible to process the images and scan for candidates in the field, we found this was more easily accomplished out of the field. Our system of deployment was to acquire images for a full day and process the images during the evening, when potential meteorite candidates were flagged for follow-up during a second trip to the field site the next day.

## 3 Field Test and Results

### 3.1 Potential meteorite fall at Walker Lake, NV

On July 14, 2019, at 09:36:50 UTC, a bright meteor of 4.1 second duration was captured by four stations of the NASA Meteorite Tracking and Recovery Network, part of the Global Fireball Observatory. The fireball track was detected from stations at Lick Observatory, Mt. Umunhum, Sunnyvale, and the Allen Telescope Array (Figure 2). The images were calibrated and the meteor track extracted (Devillepoix et al., 2020). The beginning height of the meteor was at 82.054 km and the end height of 27.066 km over the Sierra Nevada mountains. The approach orbit was asteroidal, with a = 2.358 +/- 0.032 AU and i = 10.72 +/- 0.08 degrees. The entry speed was relatively high at 21.937 +/- 0.114 km/s, but the surviving mass was calculated to be as much as 35.3 +/- 3.7 kg (EKS model (Devillepoix et al., 2020)). The meteor had a 41° slope with lateral uncertainty in the trajectory of order 250-m due to a relatively large minimum distance of 275 km between the meteor and the stations.

Analysis of the wind drift during the dark flight calculations were done using the Oakland Wind sonde data before and after the time of fall, using the software WIND with an assumed density of 3.2 g/cm$^3$. This resulted in expected fragments of 1g, 10g, 100g, and 1 kg located near Walker Lake, Nevada within the Walker River Indian Reservation of the Walker River Paiute Tribe, headquartered at Schurz, NV. The expected strewn field location is shown in Figure 3. Solutions were calculated starting falling from the final flare (white markers and from the final observed point of the trajectory (blue). Winds in this case were such that small meteorites were blown towards the larger ones, shortening the strewn field. The area is a former lake bed of Walker Lake and has sand dunes and stretches of rock-strewn flat areas. The vegetation is minimal, but increases in density towards the river.

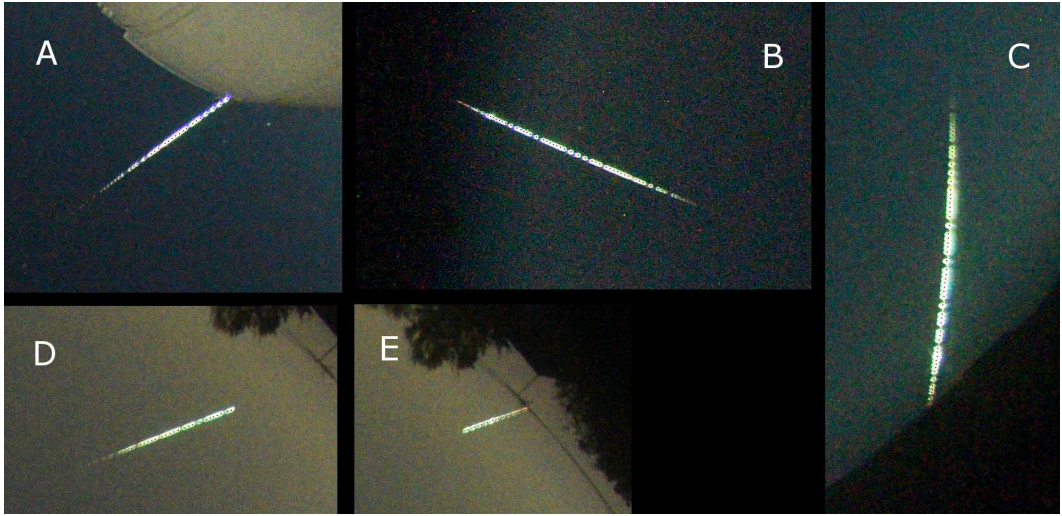

Figure 2: NASA Meteorite Tracking and Recovery imaging of the 2019 July 14 bolide from A: Lick Observatory, B: Mount Umunhum, C: Alan Telescope Array, D: Sunnyvale - early part, E: Sunnyvale -late part of bolide.

## 3.2 Field test

We received permission to conducted a field test of our meteorite classification system in Walker Lake, Nevada, the site of a meteorite fall on July 14, 2019 (Figure 3). Overall we conducted 10 test flights at two different sites within the expected strewn field. Site A was near the 10g marker in Figure 3, where we expected 1-20g fragments. Site B was near the 100g marker in Figure 3, where we expected 50-150g fragments. Due to the smaller expected fragment size at Site A, we flew the drown at a lower altitude (2-3 m). At Site A we also deployed our collection of 8 meteorite fragments during each test flight to confirm our ability to detect the meteorites used to train our classifier on a new terrain type. At Site B we conducted higher altitude (3-6 m) surveys to cover a larger area and potentially find a single large fragment $\sim$ 100 g.

A summary of the 10 test flights and results is shown in Table 1. Four tests were conducted at Site A, three at 2 m and one at 3 m. Six tests flights were conducted at Site B, two at 3 m and four at 6 m. During each test flight the drone acquired images for $\sim$ 20 minutes, limited by the battery life of the drone. Each test flight acquired 129-388 full GoPro images. These full 4000x3000 pixel images were spliced into overlapping image patches 1000x600 pixels. The image patches were analyzed with the trained RetinaNet model. An image patch was classified as 'positive' if it contained an object classified as a meteorite with a score of 0.5 or higher, and all other image patches were classified as 'negative'. The number of positive and negative image patches is listed in Table 1. Each test flight resulted in a large number of image patches classified as positive. This was particularly true for the low-altitude flights, where there were on average $\sim$ 2650 positive patches per test flight ($\sim$ 28% of the all patches). This is mostly due to the large number of other rocks in the survey area, and illustrates the difficulty of locating meteorite remotely on rocky terrain. The higher altitude flights (6 m) resulted in less positive patches, with an average of $\sim$ 1112 positive patches per test flight ($\sim$ 12% of the all patches). Test Flight 5 contained around three times as many positive patches as the other 6 m flights, which was likely due to the more rocky terrain in the Test 5 flight path.

The results illustrated in Table 1 show that RetinaNet returns a large number of false positives for a given search area. For the 2−3 m Test Flights, the number of positive patches returned was prohibitive, meaning that the thousands of positive patches flagged for follow-up is too cumbersome to sort through and identify expected meteorite fragments. For the 6 m test flights, the number of positive patches was sufficiently low to allow a user to quickly scan through the positive image patches and identify fragments to follow-up with an in-field visit.

Apart from generally determining the number of positive patches during field surveys, we conducted two critical tests to determine the feasibility of using our system in the field. The first test was to deploy our collection of known meteorite fragments in the drone survey path and test the ability of our machine learning algorithm to correctly identify these fragments on novel terrain. The second test was to scan through the most likely meteorite candidates flagged by our trained RetinaNet model and attempt to relocate them in the field based on the drone-acquired images and geolocation determined from cross-referencing the image timestamp and the drone GPS log.

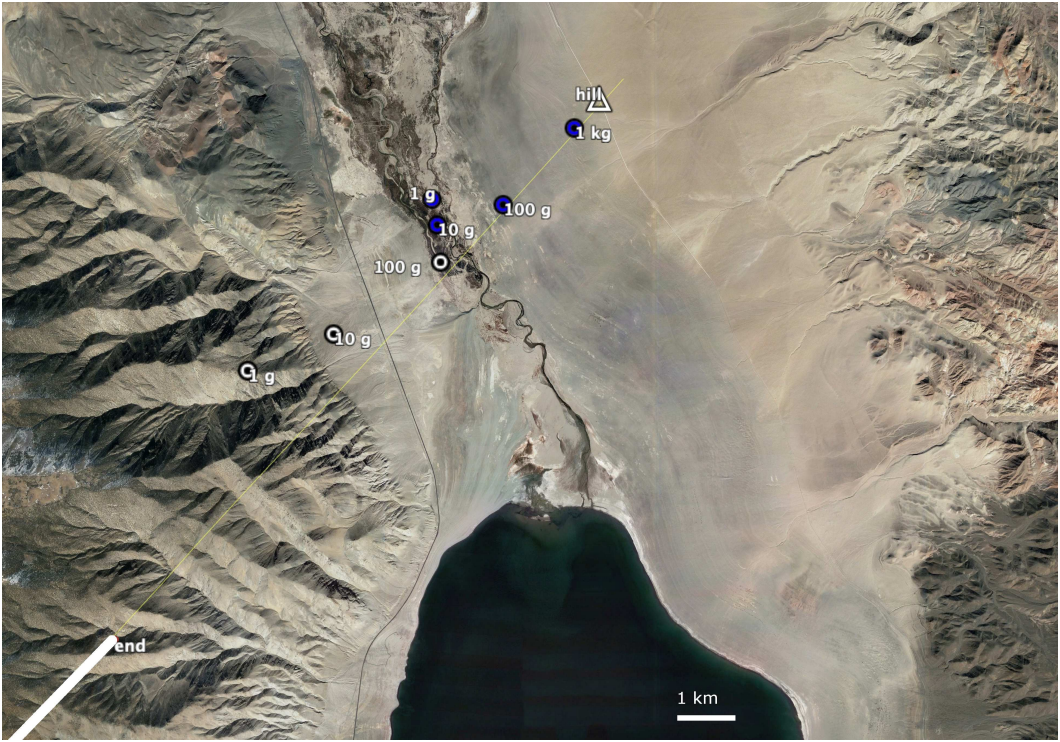

Figure 3: Map of the expected Walker Lake strewn field. The while line indicates the final part of the meteor trajectory based on NASA Meteorite Tracking and Reocvery imagery. The yellow line is an extension. The dots mark the anticipated location for meteorites falling from the final flare (white markers) and the end point (blue markers), based on Oakland wind sonde data and the WIND darkflight model.

Table 1: Field test runs

| Test | Site | Altitude (m) | Images | Total patches | Positives | Negatives |
|---|---|---|---|---|---|---|
| 1 | A | 2 | 256 | 12240 | 3553 | 8687 |
| 2 | A | 2 | 245 | 11232 | 2322 | 8910 |
| 3 | A | 2 | 288 | 13824 | 1328 | 12496 |
| 4 | A | 3 | 145 | 6960 | 679 | 6281 |
| 5 | B | 3 | 135 | 6480 | 2019 | 4461 |
| 6 | B | 3 | 129 | 6192 | 2769 | 3423 |
| 7 | B | 6 | 214 | 10272 | 2377 | 7895 |
| 8 | B | 6 | 175 | 8400 | 875 | 7525 |
| 9 | B | 6 | 190 | 9120 | 399 | 8721 |
| 10 | B | 6 | 225 | 10800 | 800 | 10000 |

To test the ability of our trained object detection network to correctly identify our collection of known meteorite fragments on novel terrain, during each test survey at Site A we deployed several of the 8 meteorite samples in our collection in the drone flight path. An example image from a test survey in which the drone flew over the deployed meteorites is shown in Figure 4. As described in Section 2, each full GoPro image was split into smaller 1000x600 pixel image patches that were run through the object detection model. Examples of image patches containing correctly identified meteorites and false positives are shown in Figure 5. The two meteorites from Figure 4 were correctly identified in the smaller image patches and given a model score 1.0 by the object detection algorithm (Figure 5a). Several false positives were also identified with two examples shown Figure 5b, which were given a score of 0.89 and 0.76. Overall, of the four low-altitude test flights where we deployed meteorite fragments in the survey path, our object detection model correctly located all of the deployed meteorite fragments. This is not surprising because the meteorite fragments we deployed in the field were the same meteorite fragments used to train our model. However, we did deploy the meteorites on completely new terrain than was used during the model training, illustrating the ability of our object detection network to locate our meteorite samples on a new terrain not used in the training data.

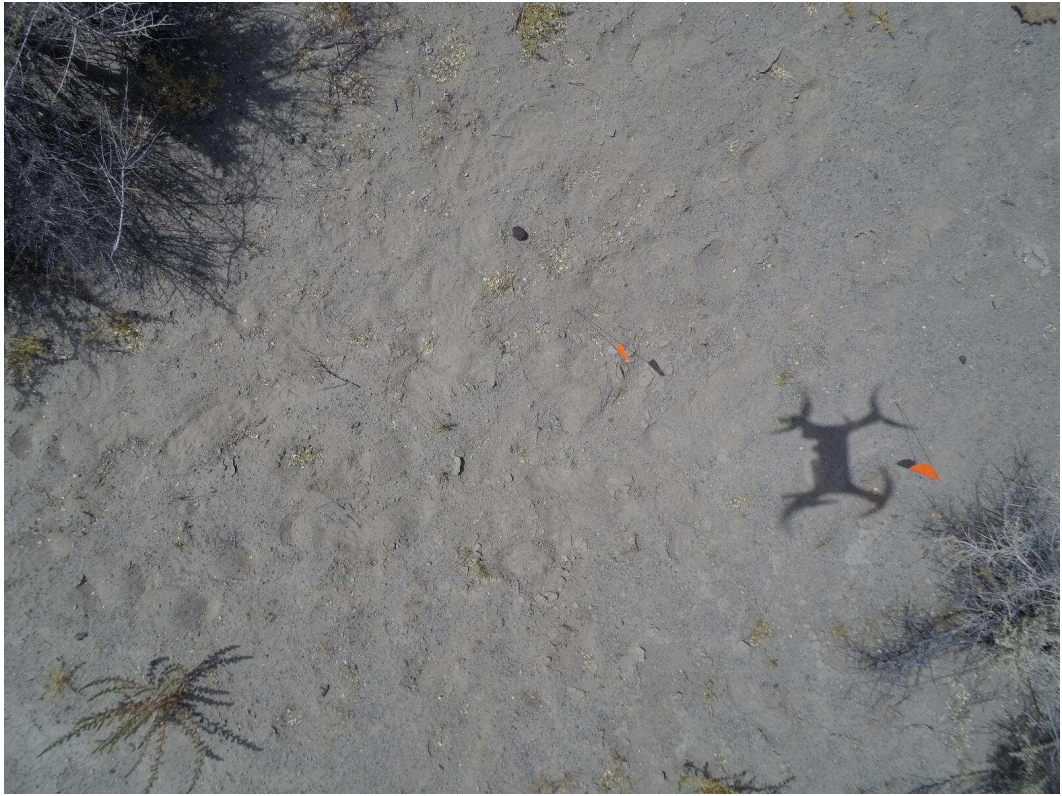

Figure 4: Example image of two meteorites deployed during a field test near Walker Lake, Nevada. The meteorites are marked with orange flags.

To test our ability to locate candidate meteorite flagged by our object detection algorithm in the field, we scanned through the positive image patches from the higher altitude (6 m) surveys conducted at Site B. We identified three meteorite candidates to attempt to locate in the field. The three drone images flagged for follow-up are shown in Figure 6 and Supplementary Figures A.1 and A.2. The GPS coordinates of the three meteorite candidates were referenced to the drone flight log to determine the expected GPS coordinates where the images were acquired. During the a second trip to the field the day after the flight tests, we visited the expected GPS coordinates of each of the three candidate meteorites and used the drone images to locate the meteorites. We successfully located the three meteorite candidates. Close up images of the three meteorite candidates located in the field are shown in Figure 7. Only two of the three stones we had flagged appeared a good meteorite candidate upon closer in-situ examination (Figure 7 a and b). These two meteorite were dark rocks, similar in appearance to meteorites. The candidate that was not a good match to a meteorite fragment (Figure 7 c, corresponding to Supplementary Figure A.2) was on closer examination a dark brush combined with a shadow. This illustrates that in some cases shadows, poor resolution, and camera perspective can confuse the identification of meteorite candidates. Still, two candidates showed a good match to what might be expected from a fresh meteorite fall and were able to be located in the field after georeferencing the flagged images. As a proof of concept, this illustrated our ability to locate in the field prospective meteorites that were flagged in images acquired from prior field surveys.

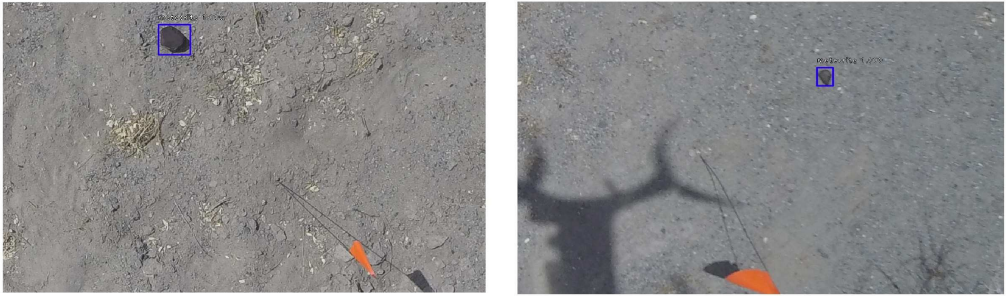
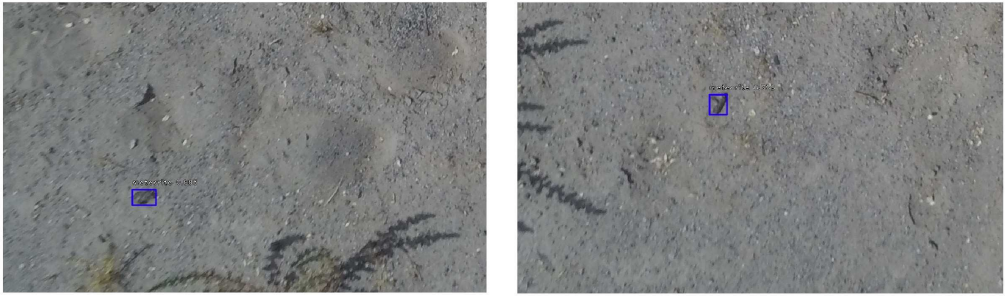

Figure 5: Example 1000x600 pixel image patches from Figure 4 containing (a) correctly identified meteorites (score = 1.0) that we placed in the test area from our collection and (b) false positives (scores = 0.89 and 0.76).

## 4 Discussion

Our field deployment highlighted several areas of improvement in order to make drone searches for meteorite fragments fully effective. Our focus and the primary area of improvement is the object detection classifier. Because our local collection of meteor fragments was limited to 8 samples, with the exception of the training data acquired from the web our training data was biased towards locating these 8 samples. Expanding the training dataset to include more meteorite images from the web and more drone acquired images of meteorites other than the 8 we had in our collection would increase the diversity of meteorites in the training data. Every time the drone system is deployed in the field it generates new data that can later be added to the training data to create a newly trained object detection model. Bringing new meteorites not previously in our collection and deploying them during field tests would allow for continued improvement of our trained object detection model, however, this option is limited by the availability of such meteorites. We also noticed during our tests that the accuracy of the classifier was related to several variables, including the resolution of the training and test images. It is possible that an improved camera resolution or more stable height control could decrease the amount of false positives.

In addition to updating the training dataset, the object detection network and method of training the final model could also be improved. For example, we could determine if certain iterations of training data result in more efficient object detection by training RetinaNet with varied training data sets, adjusting factors such as the ratio of images acquired from the web to images of our collection of 8 fragments, or changing the image resolution or size of the training data. We could also test other object detection networks to determine if certain networks are more efficient when trained on our dataset. Although RetinaNet represents the state of the art in machine learning and is widely used in computer vision and object detection, machine learning is a rapidly evolving field and it is possible that a newer machine learning approach could be even more effective. Future work could more directly compare the effectiveness of various types of object detection networks.

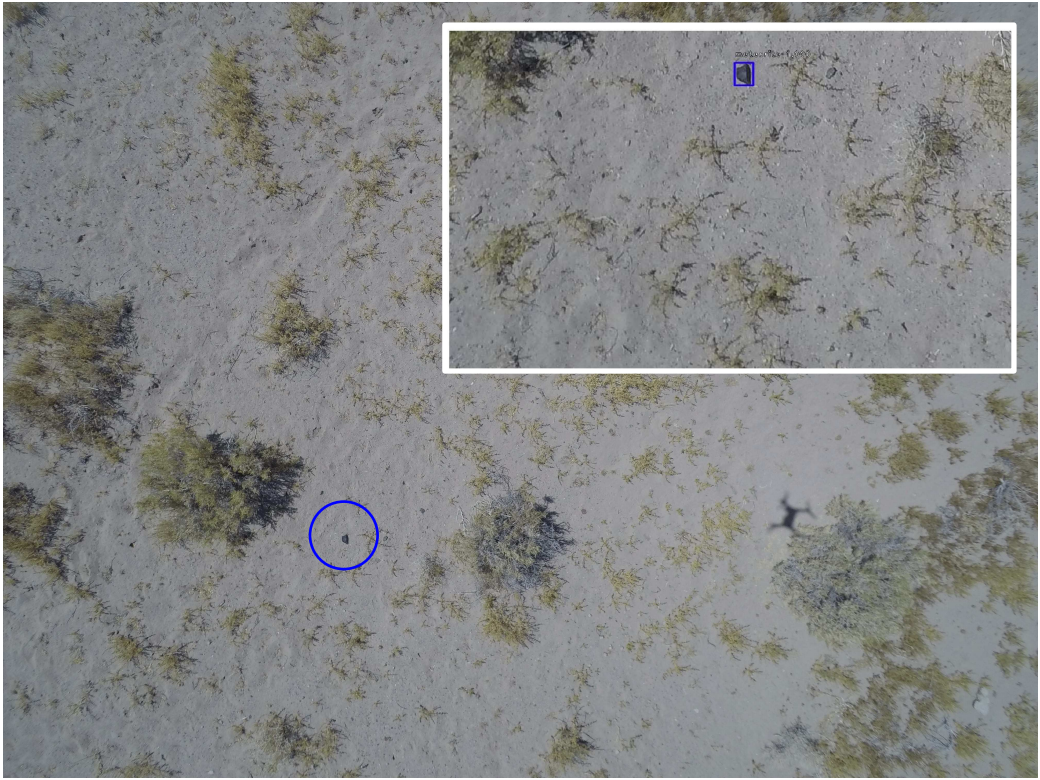

Figure 6: Example image showing a candidate meteorite identified during the higher altitude field test flights. The large image shows the full 4000x3000 pixel GoPro image obtained from the drone, with the candidate meteorite circled in blue. The inset shows the enlarged 1000x600 pixel image patch run through the object detection model with the meteorite outlined in a blue box (model likelihood = 1.0).

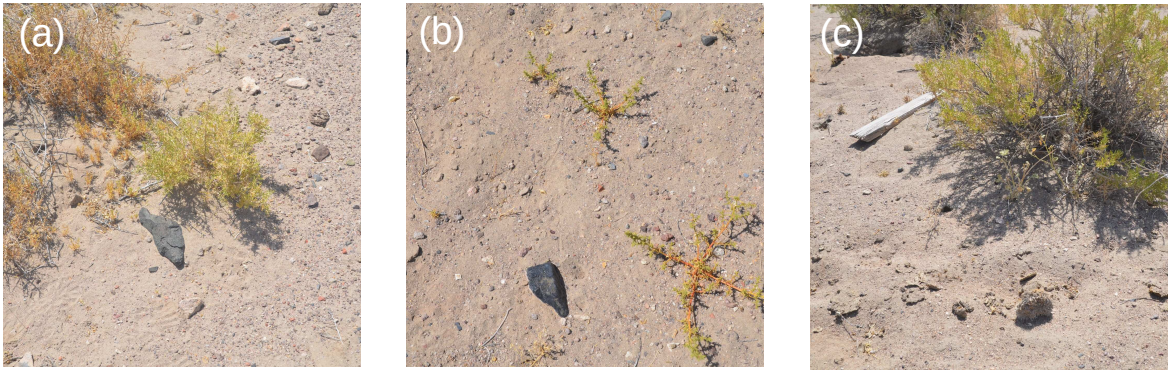

Figure 7: Images of meteorite candidates taken close-up with a DSLR camera after being geolocated in the field. Panels (a), (b), and (c) correspond to candidates in the drone imagery identified in Figure 6 and Supplementary Figures A.1 and A.2, respectively.

In addition to improvements in the machine learning algorithm, our field test identified some necessary hardware improvements. We found that surveys at 2 or 3 m covered too small of an area to be particularly effective. These surveys also located more false positives than the 6 m surveys. However, the GoPro camera was not at high enough resolution to be fully effective for surveys conducted at 6 m altitude. Camera wobble also contributed to minor image blur which made object detection more difficult. To improve our system requires a drone with an upgraded camera that can take higher resolution photos from 6 m. This is not possible with the 3DR Solo Drone, which can only mount a GoPro camera, so a new drone system is required. Unfortunately, there is no economical commercial drone that can use a laser altimeter to maintain constant elevation above the ground, which is necessary to obtain a relatively uniform image resolution. However, drone technology is continually being improved, so a new hardware solution that is within reach of regular consumers may present itself in the future.

The original intention of the project was to process the images on-site with the object detection classifier so flagged positives could be immediately checked during the same trip to the field. However, our field test found this to be ineffective. Images were difficult to scan through in the field on the laptop due to the outdoor lighting conditions. The relatively fast image processing time of $\sim 6$ s means that in theory images could be processed at the same rate that they are acquired with drone flights. However this requires a larger team and more equipment (such as a power source for the laptop) so that one user could run the classifier on the images while another user conducts the next test flight.

Overall, the object detection network deployed in this work presented a major improvement over the direct image classifier used in our previous field test (Citron et al., 2017). Our results showed that the trained RetinaNet object detection network was highly effective on our dataset, and could locate meteorites on terrain not used for training the model. RetinaNet reduced the image processing time and increased the accuracy of our model. While the amount of false positives were still high, for 6 m flights there was a sufficiently low number of positive image patches that a user could scan through the data in a reasonable timeframe and flag meteorite candidates for follow-up study. The ability to locate flagged meteorites in subsequent field excursions demonstrates the system can be used to locate freshly fallen meteorite fragments.

## 5 Conclusions

We demonstrated that as a proof of concept, it is possible to identify meteorites in the field by applying an object detection classifier to images acquired with an autonomous drone. As a first step at automating meteorite detection, we constructed a large dataset of meteorite images and trained an object detection network. We demonstrated the accuracy of our classifier in the field over terrain not used in our training dataset, and were able to locate meteorite candidates identified by our classifier during subsequent field trips. With a larger training dataset, updated classification scheme, and improved imaging hardware, machine learning coupled to an autonomous drone survey could prove a valuable tool for increasing the number of meteorite fragments found from fresh falls. This is particularly important for fresh meteorite falls where only small fragments are expected. These falls are unlikely to draw the attention of meteorite hunters, but if the fall is imaged with an all-sky survey, locating fragments is essential to augmenting the number of freshly fallen meteorites with entry orbits computed from imaged fireball trajectories. By constructing a more efficient classifier for locating meteorites in the field, it may be possible to increase the number of meteorites found that can be associated with imaged fireball trajectories, increasing our understanding of the composition of meteors and their parent asteroid bodies.

## Acknowledgements

This project was completed as part of the 2016 NASA Frontier Development Lab, which was supported by the NASA Office of the Chief Technologist. We would like to thank Jason Utas for assistance with our prior Creston, CA field study. We also thank Fabio Teixeira and Brian Lim of Hypercube and Carlos Uranga for help testing various drone designs. We thank James Parr and Jordan McRae, co-directors of FDL, as well as Jim Adams, deputy chief technologist for NASA, Bruce Pittman, Chief Systems Engineer at NASA Ames Research Center, Bill Diamond, CEO of the SETI Institute, Debbie Kolyer, Grants Manager of the SETI Institute, Jonathan Knowles, explorer in Residence at Autodesk, Alison Lowndes, Deep Learning Solutions for NVIDIA, Eric Dahlstrom, President of the International Space University, Yarin Gal, University of Cambridge Research Fellow. The field work was coordinated with the tribal leaders of the Walker River Northern Paiute Tribe. We thank especially Cheri Clenneding and her interns for assisting in the search.

The Global Fireball Observatory is supported by the Australian Research Council. RC and PJ are supported by NASA grant 80NSSC18K08.

# A  Supplementary Figures

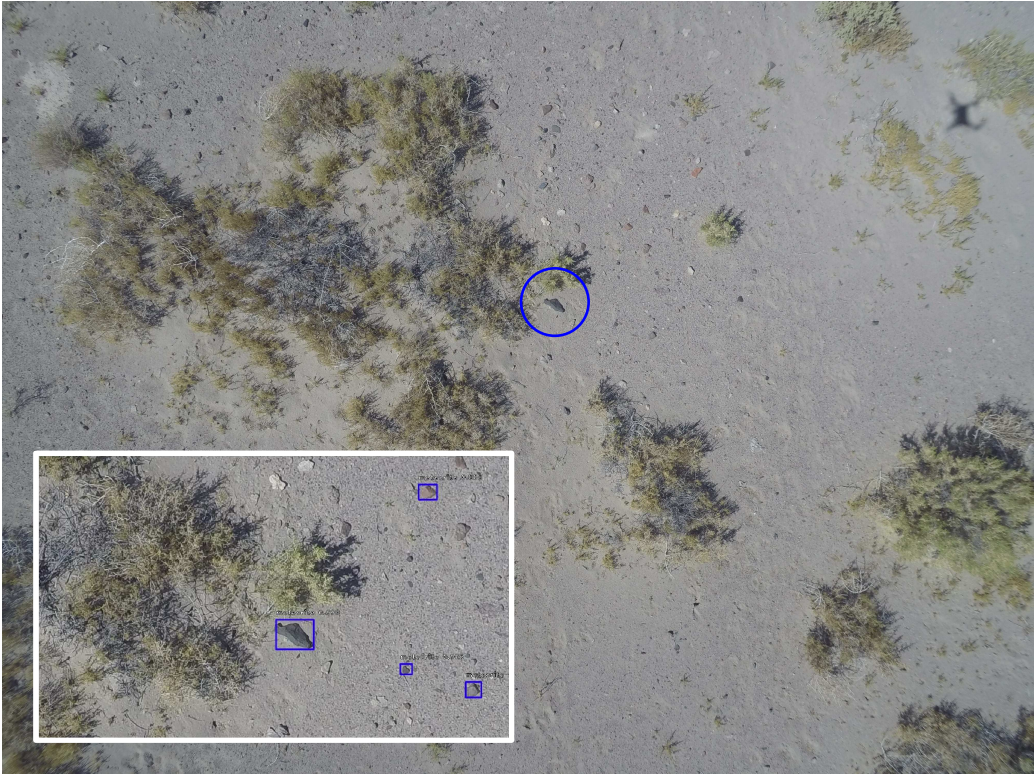

Figure A.1: Example image showing a candidate meteorite identified during the higher altitude field test flights. The large image shows the full 4000x3000 pixel GoPro image obtained from the drone, with the candidate meteorite circled in blue. The inset shows the enlarged 1000x600 pixel image patch run through the object detection model with the meteorite outlined in a blue box (model likelihood = 0.992).

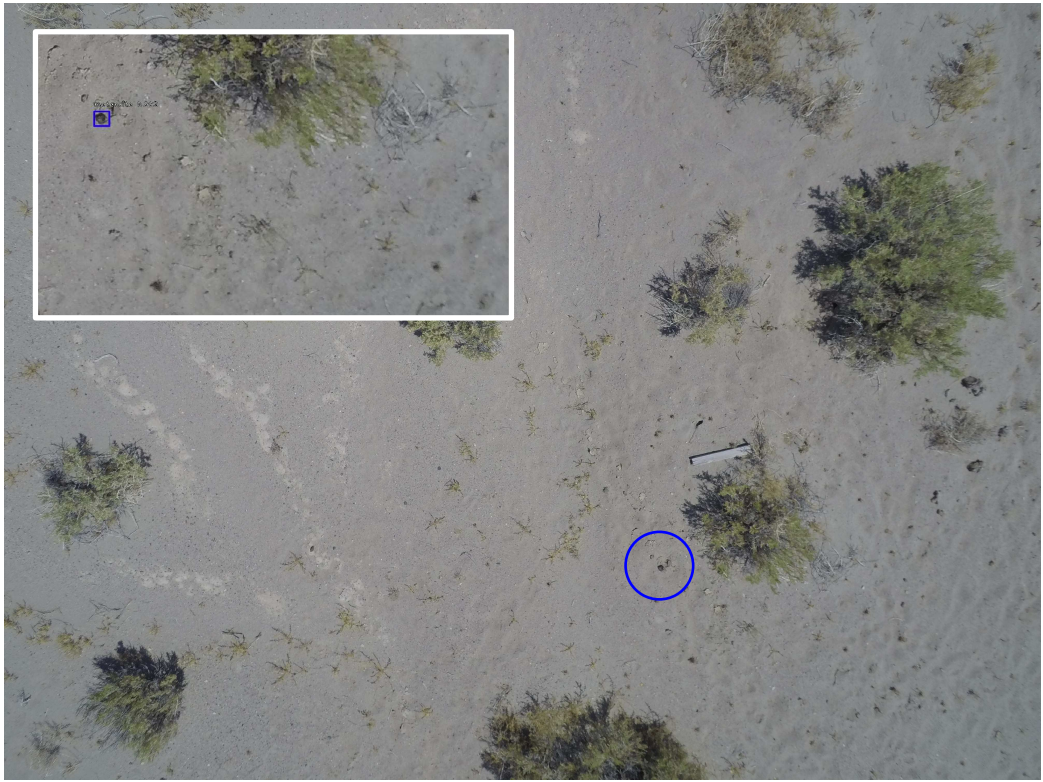

Figure A.2: Example image showing a candidate meteorite identified during the higher altitude field test flights. The large image shows the full 4000x3000 pixel GoPro image obtained from the drone, with the candidate meteorite circled in blue. The inset shows the enlarged 1000x600 pixel image patch run through the object detection model with the meteorite outlined in a blue box (model likelihood = 0.992).